\documentclass{aastex631}

\usepackage{graphicx}
\usepackage{tikz}
\usetikzlibrary{matrix}
\usepackage{rotating}
\usepackage{afterpage}
\usepackage{enumitem}
\usepackage{float}

\begin{document}

\title{Reevaluating the Origin of Detectable Cataclysmic Variables in Globular Clusters: Testing the importance of Dynamics}

\author{Liliana Rivera Sandoval}
\affiliation{Department of Physics and Astronomy, University of Texas Rio Grande Valley, Brownsville, TX 78520, USA}

\author{Diogo Belloni}
\affiliation{Departamento de F\'isica, Universidad T\'ecnica Federico Santa Mar\'ia, Av. España 1680, Valpara\'iso, Chile}

\author{Miriam Ramos Arevalo}
\affiliation{Department of Physics and Astronomy, University of Texas Rio Grande Valley, Brownsville, TX 78520, USA}

\begin{abstract}
Based on the current detectable cataclysmic variable (CV) population in Galactic globular clusters (GCs), we show that there is not a clear relation between the number of sources per unit of mass and the stellar encounter rate, the cluster mass, or the cluster central density. If any, only in the case of core-collapsed GCs could there be an anticorrelation with the stellar encounter rate.
Our findings contrast with previous studies where clear positive correlations were identified. 
Our results suggest that correlations between faint X-ray sources, from which often conclusions for the CV population are drawn, and the GC parameters considered here, are likely influenced by other type of X-ray sources, including other types of compact binaries, which have X-ray luminosities similar to CVs.
The findings presented here also suggest that the role of primordial systems is more important than previously believed and that dynamical formation has less influence in the current detectable CV population. The long-standing paradigm that GCs are efficient factories of CVs formed via dynamical interactions does not seem to be supported by current observations.  
\end{abstract}


\section{Introduction} \label{sec:intro}

Cataclysmic variables (CVs) are binaries in which a white dwarf (WD) accretes matter from a main-sequence (MS) star. They typically have orbital periods between 80 minutes and 10 hr \citep{2023Inight}.
Due to crowding, the identification of CVs in globular clusters (GCs) is challenging. Yet, they have been confirmed spectroscopically and mainly studied photometrically in several Galactic GCs (for a review of CVs in GCs, see \cite{2021belloni}).\looseness=-10

CVs in GCs are usually first identified by their X-ray emission, i.e., luminosities and spectrum characteristics. Their X-ray properties are strongly influenced by the magnetic field intensity of the accretor. Nonmagnetic CVs have typical X-ray luminosities $L_X<10^{32}$ erg s$^{-1}$, while the magnetic ones, such as intermediate polars (IPs) can have $L_X >10^{33}$ erg s$^{-1}$ \citep{2020Bahramian}. However, IPs seem to be scarce in GCs, possibly due to 
the rotation- and crystallization-driven dynamo scenario \citep{2021bBelloni}. The most suitable model for characterizing the spectra of CVs is a thermal plasma model, which includes a Bremsstrahlung continuum along with emission lines.

Once potential systems have been identified in X-rays, their counterpart is later located in optical and/or ultraviolet (UV) within the X-ray error circle. This provides a more confident classification and further characterization. In optical/UV, CVs are expected to be bluer than normal MS stars in the color-magnitude diagrams (CMDs) of GCs given the contribution of the accreting WD and the accretion disk. Because 
the donors in CVs are hydrogen-rich, one would also expect to see H$\alpha$ line emission coming from the heating and ionization of hydrogen in the accretion disk. Therefore, in a CMD that includes a narrow filter that encloses the H$\alpha$ line, one would see a H$\alpha$ excess when compared to MS stars of the same magnitude. Furthermore, variability in one or more filters is also a signature of their binary interaction and/or presence of a disk. A combination of these techniques is what has allowed, in the last decades, a more confident classification of faint X-ray sources ($<10^{34}$ erg s$^{-1}$), which include a large population of chromospherically active binaries (ABs), CVs, background/foreground objects, quiescent low-mass X-ray binaries (qLMXBs) and millisecond pulsars\footnote{MSPs are rapidly rotating neutron stars. LMXBs are binaries with a neutron star or a black hole accreting from a low-mass star. In a quiescent state, the mass transfer and accretion onto the compact object are minimal, resulting in a reduction in X-ray emission. The identification of MSPs and qLMXBs in GCs has also benefited from radio studies.} (MSPs), compared to only X-rays studies of CVs in GCs.

The Chandra X-ray Observatory, together with the Hubble Space telescope (HST) have been key in the
classification of X-ray sources given their very good spatial resolution and sensitivity, needed in very crowded environments such as GCs. So far, HST is the only telescope that resolves the cores of GCs in the UV and optical, where many CV systems reside. Reaching very deep observations, the HST has allowed the detection of very faint sources, which are difficult to detect from the ground or with other space telescopes. In particular, two instruments on HST that have been pivotal for the further classification of Chandra X-ray sources are the Wide Field Camera 3 (WFC3) and the Advanced Camera for Surveys (ACS). The WFC3 can obtain images in the near ultraviolet (NUV) with excellent spatial resolution, while the ACS is more sensitive in the optical (particularly convenient when using narrow filters such as H$\alpha$) and has instruments that can obtain information in the far-UV (FUV), thus improving the CV classification. To date the largest samples of CVs have been obtained for the GCs 47 Tuc, $\omega$ Cen, NGC 6397, and NGC 6752 (See Appendix \ref{appendix_ref}). This is because these are clusters with low reddening, which allows UV and deep optical searches; they are relatively close by and are massive clusters that harbor large amounts of compact binaries that emit X-rays, and hence more CVs than other clusters. While the observed samples are not complete in those clusters (with respect to the total expected CV population), the very deep observations obtained with telescopes like HST and Chandra have led to meaningful samples as very faint sources have been detected allowing a better comparison to models.

For decades, the formation of CVs in GCs has been thought to be largely influenced by dynamical interactions \citep[e.g.,][]{2003Pooley,2006Pooley,2006Ivanova,2006Shara,2017Hong}, with exchanges playing a very important role. One way to assess the importance of these interactions is through the present-day stellar encounter rate $\Gamma$. 
This parameter is considered an effective way to measure the effect of stellar interactions in GCs because it provides a quantitative measure of how frequently individual stars in the cluster come close enough to interact gravitationally with one another\footnote{We shall mention that, as already highlighted by \citet{2021belloni}, $\Gamma$ depends on GC properties that might change quite substantially over the cluster lifetime, and for this reason, present-day $\Gamma$ is most likely reliable to address the impact of dynamics on the formation of stellar populations only in the past few Gyr.}. It is calculated based on several parameters, such as the central density ($\rho_c$), the core radius, and the central velocity dispersion, instead of a single GC parameter. If dynamical interactions have a relevant impact on the formation of CVs, one might then anticipate observing a correlation between the number of detected CVs within a cluster and $\Gamma$. Several studies have been carried out in this regard, using faint X-ray sources ($<10^{34}$ erg s$^{-1}$) under the assumption that CVs substantially contribute to the X-ray emission of that population. One of the first studies was carried out by \cite{2006Pooley} who found a correlation between the specific encounter rate $\gamma$ ($\Gamma / M$ where $M$ is the cluster mass) and the specific number of sources $n_{II}$ ($N_{II}/M$, with $N_{II}$ the number of faint X-ray sources per cluster that the authors considered to be mostly CVs and ABs). 
They found a positive power-law correlation and concluded that most faint X-ray sources, i.e., CVs and ABs in dense GCs, were formed via dynamical interactions. Since then, there has been a debate on the role of dynamics in the formation of CVs in GCs. 

In this document we investigate the relation between the specific number of detectable CVs and the specific stellar encounter rate, the central density and the cluster mass using
the information available of CVs in Galactic GCs in the literature, to assess the importance of dynamics in the population of these binaries. Here ``detectable" means the present-day population of CVs, which have X-ray luminosity $> 5 \times10^{29}$ erg s$^{-1}$ like in \cite{2019Belloni} and/or which is bright enough to be detectable in the optical and/or UV.

\begin{table}
\begin{center}
\begin{tabular}{ l c c c c c c c c}
\hline
 Name  & RA & DEC & Distance$^1$ & Mass$^1$ & $\Gamma$ $^2$ & $\log \rho_{c}^1$   & Lower  &Identified \\
& (J2000)& (J2000) & (kpc) & (M$_{\odot}$) &  & [M$_\odot/pc^3$]  & Lim.~L$_X$ & detectable \\
& & & &  &  & & (erg~s$^{-1}$)  & CVs \\
\hline
\multicolumn{9}{c}{Non-CC GCs}\\
\hline
NGC 104 (47~Tuc) & 00:24:05.67 & -72:04:52.6 & 4.41 & $7.79 \pm 0.05\times 10^5$ & $1000_{-134}^{+150}$ & 5.31 & $2.00\times 10^{29^a}$ & 43 \\
NGC 288   & 00:52:45.24 & -26:34:57.4 & 9.80 & $1.16 \pm 0.03 \times 10^5$ & $0.766_{-0.204}^{+0.284}$& 1.62  & $7.00\times 10^{30^b}$ & 1\\
NGC 2808 & 09:12:03.10 &	 -64:51:48.6 & 9.80	& $7.42 \pm 0.05 \times 10^5$ & $923_{-82.7}^{+67.2}$	& 4.60 & $1.36 \times 10^{31}$ & 1\\
NGC 5139 ($\omega$ Cen) & 13:26:47.24 & -47:28:46.5 & 5.20 & $3.55 \pm 0.03 \times 10^6$ & $90.4_{-20.4}^{+26.6}$ & 3.22  & $1.00\times 10^{30 ^c}$ & 23 \\
NGC 5272 (M3) & 13:42:11.62 & +28:22:38.2 & 10.06 & $3.94 \pm 0.23 \times 10^5$  & $194_{-18.0}^{+33}$ & 3.80  & $1.10\times 10^{31 ^d}$ &  3 \\
NGC 5904 (M5) & 15:18:33.22  & +02:04:51.7 & 7.50 &  $3.72 \pm 0.06 \times 10^5$ & $164_{-30}^{+39}$ & 3.66 & $5.26\times 10^{30}$& 1\\
NGC 6093 (M80) & 16:17:02.41 & -22:58:33.9 & 10.50 & $2.49 \pm 0.12 \times 10^5$ &  $532_{-69}^{+59}$ & 5.48  & $7.00\times 10^{30^e}$ &  4\\
NGC 6121 (M4) & 16:23:35.22 & -26:31:32.7 & 2.14 &  $9.69 \pm 0.26 \times 10^4$  & $26.9_{-9.6}^{+11.6}$ & 3.63  & $3.10\times 10^{29}$ & 4 \\
NGC 6205 (M13) & 16:41:41.24  & +36:27:35.5 & 6.60 & $4.53 \pm 0.34 \times 10^5$  & $68.9_{-14.6}^{+18.1}$ & 3.58  & $4.82\times 10^{30}$ & 1\\
NGC 6218 (M12) & 16:47:14.18 & -01:56:54.7 & 5.22 & $8.65 \pm 0.60 \times 10^4$ & $13.0_{-4.03}^{+5.4}$ & 4.05 & $6.02\times 10^{30}$ & 1\\
NGC 6388  & 17:36:17.23 & -44:44:07.8 & 11.00 & $1.06 \pm 0.01 \times 10^6$  & $899_{-213}^{+241}$ & 5.55  & $5.00\times 10^{30^f}$ &5 \\
NGC 6402 (M14) & 17:37:36.10  & -03:14:45.3  & 9.30 &  $7.74 \pm 0.61 \times 10^5$ & $124_{-30.2}^{+31.8}$ & 3.32 & $6.45\times 10^{31}$& 1\\
NGC 6656 (M22) & 18:36:23.94 & -23:54:17.1 & 3.10 & $4.16 \pm 0.05 \times 10^5$ & $77.5_{-25.9}^{+31.5}$ & 3.66  & $8.26\times 10^{29}$ & 3\\
NGC 6809 (M55) & 19:39:59.71 & -30:57:53.1 & 5.75 & $1.88 \pm 0.12 \times 10^5$ & $3.23_{-1.0}^{+1.38}$ & 2.45 & $5.15\times 10^{30}$ & 4 \\
NGC 6838 (M71) & 19:53:46.49 & +18:46:45.1 & 3.86 & $4.91 \pm 0.47 \times 10^4$  & $1.47_{-0.138}^{+0.146}$ & 3.67  & $1.50\times 10^{30^g}$ & 1  \\

\hline
\multicolumn{9}{c}{CC GCs}\\
\hline
NGC 6397  & 17:40:42.09 & -53:40:27.6 & 2.48 & $8.89 \pm 0.16 \times 10^4$& $84.1_{-18.3}^{+17.9}$ & 5.99 &  $9.76\times 10^{28}$ & 15 \\
NGC 6624 & 18:23:40.51 & -30:21:39.7 & 7.00 & $7.31 \pm 0.20 \times 10^4$ &  $1150_{-178}^{+110}$ & 5.21  & - & 2\\
NGC 6681 (M70) & 18:43:12.76 & -32:17:31.6 & 9.20 & $1.13 \pm 0.02 \times 10^5$ & $1040_{-192}^{+270}$ & 7.33  & - &1 \\
NGC 6752  & 19:10:52.11 & -59:59:04.4 & 4.30 & $2.39 \pm 0.04 \times 10^5$  & $401_{-126}^{+182}$ & 5.38  & $3.00\times 10^{29^h}$ & 17\\
NGC 7078 (M15) & 21:29:58.33 & +12:10:01.2 & 9.90 & $4.53 \pm 0.05 \times 10^5$ & $4510_{-990}^{+1360}$ & 6.60 & $6\times 10^{30^i}$ & 8\\
NGC 7099 (M30) & 21:40:22.12 & -23:10:47.5  & 8.10 & $1.33 \pm 0.08 \times 10^5$   & $324_{-81}^{+124}$ & 5.99  &  $1.57\times 10^{30}$ & 9 \\
\hline
\end{tabular}
\caption{\label{table1} Properties of the GCs considered in this work. $^1$ Taken from \cite{2018Baumgardt}.
$^2$ Taken from \cite{2013Bahramian}.
$^a$ From faintest detection in \cite{2005Heinke} in the 0.5-6 keV band.
$^b$ From \cite{2006Kong} in the 0.3-7.0 keV band.
$^c$ From \cite{2018Henleywillis} in the 0.5-6.0 keV band.
$^d$ From \cite{2019Zhao} in the 0.5-7.0 keV band, but \cite{2003Pooley} found a limit of $4\times10^{30}$ erg s$^{-1}$.
$^e$ From \cite{2003Heinke} in the 0.5-2.5 keV band.
$^f$ From \cite{2012Maxwell} in the 0.5-6.0 keV band, \cite{2013Bahramian} finds a value of $1.74\times10^{31}$ erg s$^{-1}$ in the $0.5-10$ keV range. 
$^g$ From \cite{2008Elsner} in the $0.3-8$ keV band.
$^h$ From \cite{2021Cohn} in the $0.5-7.0$ keV band. 
$^i$ From the faintest object in \cite{Hannikainen2005} in the $0.5-7$ keV band. 
We do not provide the lower limit of $L_X$ for NGC 6624 and NGC 6681 but they have been observed in X-rays by other authors. The CC or non-CC classification was taken from \cite{2010Harris}.}
\end{center}
\end{table}

\begin{figure*}
\begin{center}
\includegraphics[width=0.99\linewidth]{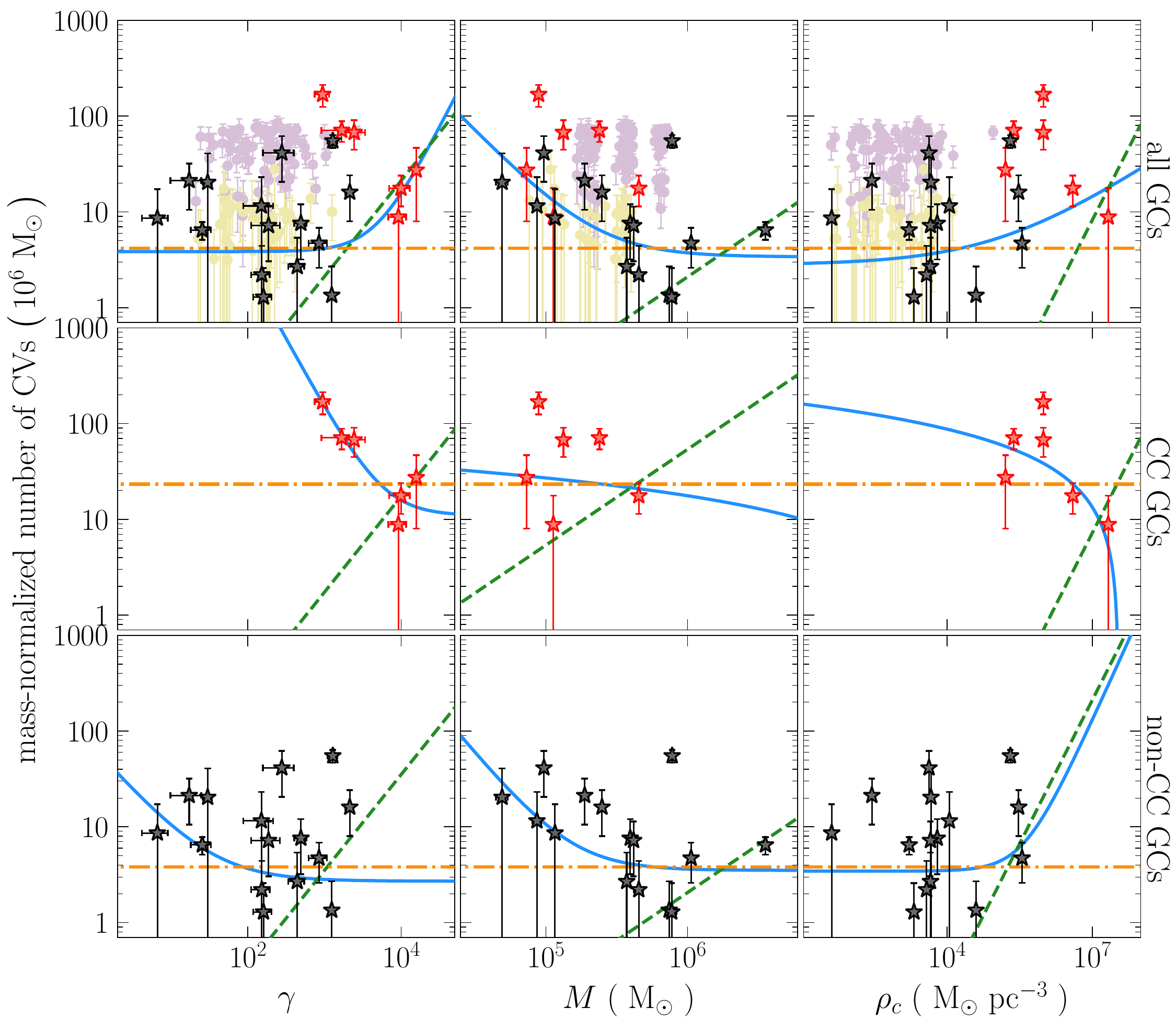}
\end{center}
\caption{Relations between $n_{\rm CV}$ and $\gamma$, cluster mass and central density for all clusters (top panels), CC GCs (middle panels) and non-CC GCs (bottom panels). The power-law fit is represented with a solid blue line, the constant fit with a dot-dashed orange line and the linear fit with a dashed green line. Purple and yellow points in the top panels refer to models by \cite{2019Belloni} using the Kroupa and standard initial binary populations, respectively. Error bars in those model points are Poisson errors.}
\label{figures_fit}
\end{figure*}

\section{Data analysis} 
\label{Data_analysis}

We have compiled data from 21 different Galactic GCs in which (mostly) Chandra X-ray sources 
have been further classified and characterized as CVs using optical and/or UV emission mainly from the HST (usually using the ACS and the WFC3). 
For many of the systems there is also associated variability in the optical and/or UV, and none has been identified as a radio source by cross-correlating with the radio catalog by \citet[][]{2020AShishkovsky}. This reduces, furthermore, the probabilities that they are MSPs or LMXBs. We summarize the properties of the GCs we considered in Table \ref{table1}. In this study we have included systems classified as ``CVs'' and ``CVs?'' in the literature. The ``CV?'' classification frequently occurs because systems are faint and therefore are difficult to observe, for example in the optical, or their counterpart is affected by brighter stars or diffraction patterns. But their classification as CV by other authors is given based on the X-rays and UV and/or optical/UV variability. Furthermore, their characteristics are not compatible with other types of close binary. We have excluded systems whose properties, according to the different authors, could also be in agreement with other interacting close binaries, for instance those which could be ABs, MSPs, qLMXBs, or active galactic nuclei. 

In order to make a fair comparison among the samples of CVs in the different clusters and reduce the contaminants as much as possible (e.g. other types of close binaries or background/foreground sources), we included only CVs that have the same identification methodology, i.e., classification based on more than one band, which is usually X-rays plus optical and/or UV. However, there are a few systems that have been classified as CVs using only optical or UV spectra. We included these spectroscopically confirmed systems as well as spectra is the unambiguous way to identify a CV. In Table \ref{table1} we did not include CVs detected only via X-rays, detected in just one single band (e.g. UV, optical, or infrared), X-ray sources 
that have more than one plausible optical/UV counterpart where one could be a CV, or whose classification is rather ambiguous (though for completeness, we listed these systems in Appendix \ref{appendix_ref}). For example, the magnetic CVs identified by \cite{2020Bahramian}; the CVs identified in the GCs NGC 6440, NGC 6266 (M62)\footnote{For example, for this cluster, there are several CV candidates for which there is the possibility of being quiescent LMXBs. Therefore, we have excluded them from our study.}, the systems in Terzan 1, Terzan 5, NGC 6626 (M28), NGC 6715 (M54), among others (see full list in the Appendix).
However, we include the cases of NGC 6624 (star A) and NGC 6656 (CV1) and NGC 6681 (we named it CV1 in this work) in which the classification was based using spectra. Several ``photometric'' CVs have also been spectroscopically confirmed, and therefore, we included them \citep[e.g., AKO9 in 47 Tuc,][]{2003Knigge}.
We consider these selection criteria are a good compromise between completeness and possible contamination of the sample. 
The CVs we chose correspond to a meaningful sample and the first attempt to make a more reliable, clean and relatively more ``homogeneous" sample based on the available results in the literature. In appendix \ref{appendix_ref} we list the names of the CVs per cluster used in this study. A detailed catalog with individual information for each CV will be provided in another work.

For the present-day encounter rate $\Gamma$ and its uncertainty, we have used the values determined by \cite{2013Bahramian} for each cluster. For all clusters we used their value determined as $\Gamma= \sigma^{-1}\int \rho^2(r) dr$ which estimated the GC luminosity density profiles by deprojecting the cluster surface brightness. This allowed \cite{2013Bahramian} to treat both core-collapsed\footnote{Core-collapse in a GC is a phenomenon in which the central region of the cluster undergoes a significant increase in stellar density, primarily because of gravitational interactions between cluster stars.} (CC) and non-core-collapsed (non-CC) clusters in a similar way, which provides consistency to our study as we are considering both types of clusters. For the distances ($D$), masses ($M$) and central densities ($\rho_c$), we used the values provided by \cite{2018Baumgardt} for all clusters. The X-ray lower luminosity limits were obtained from \cite{2020Bahramian} for several clusters, but additional references for specific clusters are given in Table \ref{table1}. 
Analogous to \cite{2006Pooley}, we define the present-day specific encounter rate as $\gamma$ = $\Gamma / M_6$ and the specific number of detectable CVs as $n_{\rm CV}=N_{\rm CV} / M_6$, where $M_6$ is the cluster mass in units of $10^6$ M$_\sun$ and $N_{\rm CV}$ is the number of detectable CVs.
We fit a function of the shape $Y = AX^{\alpha}+ C$ , where the term C represents the contribution of the primordial population and $AX^{\alpha}$ represents the dynamically formed CVs. We also fit a constant function $Y = C$ and a linear function $Y = AX$, where the term $AX$ represents the dynamical population, as in the case of the power-law fit. Note we have not used a fit of the shape $Y = AX + C$ because that is just a particular case of the power-law. \looseness=-10

\section{Results and discussion}\label{sec:discussion}

In Figure \ref{figures_fit} we show the present-day specific encounter rate, central density, and cluster mass as a function of the specific number of detectable CVs for all clusters, CC GCs, and non-CC GCs. 
The three different fits (power-law, constant, and linear) to the data are presented in each plot. In the same figure, we also plotted results for detectable CVs in GCs obtained with the MOCCA code using the Kroupa and Standard initial binary populations as described in \cite{2019Belloni}. Vertical errors in Figure \ref{figures_fit} include the Poisson and mass errors. Horizontal errors include errors for $\Gamma$ as given by \cite{2013Bahramian} and for $\rho_c$ and for $M$ as provided by \cite{2018Baumgardt}. 

Contrary to previous observational studies of CVs \citep[e.g.,][]{2003Pooley,2006Pooley,2012Maxwell}, except for one case, we did not observe any strong correlation between the parameters for any type of GC (see Table \ref{table2}). All the fits have extremely low $p$-value, which here  is understood as the likelihood of achieving the observed value of $\chi^2$ given the assumption that the model is accurate and the best-fit parameters represent the true values. Just in the case of a power-law fit in the $\gamma$ vs $n_{\rm CV}$ relation for CC GCs, we observe a $p$-value above the $0.05$ significance level, which suggests there is an anticorrelation between these parameters for CC GCs. This agrees with the interpretation in which the larger the number of stellar interactions, the larger the rate of destruction of CV progenitors. We also performed Pearson and Spearman rank correlation tests to check for a possible correlation among the data under the null hypothesis that there is no correlation between the parameters. Except for one case, we found that the Pearson and Spearman $p$-value is considerably larger than the $0.05$ significance level (including the cases with $\rho_c > 10^4\,{\rm M}_{\sun}{\rm pc}^{-3}$), indicating that there is no significant evidence against the null hypothesis. Just in the case of the relation $\gamma$ vs $n_{\rm CV}$ for CC GCs, a value slightly above $0.05$ was identified, which is consistent with our analysis based on the fitting schemes. If we remove those clusters from Table \ref{table1} that have one or two CVs and perform the same analysis, we observe anticorrelations with $\gamma$ and $M$ for CC GCs above the $0.05$ significance level.

The fact the data were not fit by a constant function suggests the detectable population of CVs does not have a pure primordial origin. On the other hand, the bad linear fit suggests that the population does not have pure dynamical origin. 
Our findings disagree with earlier studies that suggested that the brightest CVs (easier to detect and therefore considered in this study) primarily formed dynamically through exchanges \citep[e.g.,][]{2006Ivanova,2006Shara,2017Hong}. For instance, \cite{2017Hong} found a clear correlation
between $\gamma$ and $n_{\rm CV}$ when considering only CVs formed via dynamical exchanges. However, the data here presented do not support that. In fact, we also considered the situation with only optical/UV ``bright" CVs, which we define as those that have an absolute magnitude brighter than 9 in an optical band\footnote{Currently, there is no general consensus of what a bright CV is as not all GCs have a visible bright and faint population as it has been identified in CC clusters. Furthermore, not all the CVs in GCs have been observed with the exact same filters; hence here we used the term optical to refer for example, to HST filters F435W, F438W, F555W, F606W or the traditional V. An exception is M15, whose CVs have been identified only in FUV having those absolute magnitudes $M_{FUV} \ll 9$.}.
This is consistent with the definition of bright CVs used in \cite{2019Belloni}, which in turn is based on the characteristics of the bright population of CVs identified in CC GCs \citep{2010cohn,2021Cohn}. Under this assumption, many GCs in our sample just had 1 or 2 bright CVs detected. 
 When we included all the clusters in Table \ref{table1}, regardless of the number of bright CVs (Fig.
 \ref{figures_fit_bright} in appendix \ref{appendix_ref}), we observe a negative power-law correlation for CC GCs with all GC variables with significance level larger than $0.05$. For non-CC GCs, there is also a power-law anticorrelation with $M$, but a positive power-law correlation with $\gamma$ and with $\rho_c$  
 with significance level larger than $0.05$. When we remove the clusters that have 1 or 2 bright CVs, the same results hold for CC GCs, while for non-CC GCs the positive correlation with $\gamma$ and the 
 anticorrelation with $M$ remain. The findings for CC GCs suggest that dynamics does not seem to favor the production of bright CVs. Interpreting the results of non-CC GCs for the bright CVs is more challenging. One should keep in mind the very low statistics we are dealing with, plus the fact that non-CC GCs do not have a clear intrinsic bright population as it is the case for CC GCs, i.e., these bright CVs have been artificially selected just for consistency. However, it is still interesting that there is an anticorrelation with $M$ regardless of the type of cluster.

When considering all CVs in Table \ref{table1} (regardless of their absolute optical magnitude), our results agree with the findings of faint X-ray sources by \cite{2018Cheng} and detectable CVs by \cite{2019Belloni}. \cite{2018Cheng} did not find a correlation between the cluster X-ray emissivity, defined as the X-ray luminosity per unit stellar mass, and global cluster properties such as metallicity, dynamical age, and structural parameters. 

Furthermore, \cite{2018Cheng} have shown a weak correlation between the abundance of X-ray sources and the GC stellar encounter rate in a larger GC sample and using a different method compared to \cite{2006Pooley}. A less pronounced correlation was also identified by
\cite{2010Pooley} when including a significantly larger number (by approximately a factor of 3) of GCs compared to \cite{2006Pooley}. In that study, despite the fact that the correlation was not quantified, it was evident that CC GCs showed a much flatter correlation than non-CC GCs. While the results by \cite{2018Cheng} might be influenced by measurements that are less precise due to some distant clusters or clusters with a limited number of X-ray sources included in their sample \citep[see][for a deeper discussion on this]{2020Heinke,2021belloni}, it is important to note that in \cite{2006Pooley} there might be a bias due to small number statistics, which is important to consider when interpreting their correlation with $\gamma$. 

Taking into account the simulations carried out by \cite{2019Belloni}, we theoretically did not find any correlation between the number of detectable CVs and $\gamma$, mass, and $\rho_c$ (Figure \ref{figures_fit}). \cite{2020Heinke} also did not find a correlation between X-ray cluster emissivity and central densities below $10^4\,{\rm M}_{\sun} {\rm pc}^{-3}$,
but they found correlations for larger densities. We did not find correlations at densities higher than $10^4\,{\rm M}_{\sun}{\rm pc}^{-3}$ for detectable CVs in CC GCs or non-CC GCs. This suggests that the correlation found by \cite{2020Heinke} likely occurs because the faint X-ray population of the GCs in their sample might be dominated by other compact binaries, for example ABs, MSPs, and qLMXBs, in addition to being further affected by foreground/background sources. It is worth mentioning that one of the parameters used to classify X-ray sources as CVs is the X-ray/optical flux ratio, which helps distinguish CVs from quiescent LMXBs in a relatively easy way. However, 
so far there is not a clear dividing limit between CVs and ABs \citep{2023Lugger}, which can add uncertainty to the classification using that criterion. As discussed by \cite{2023Lugger}, ABs are anticipated to constitute a primarily primordial population within clusters, having undergone evolution originating from the shorter end of the binary period distribution. So if there is variation in the overall binary fraction among clusters, this implies that the size of the AB population should correlate with the mass of the cluster. However, this is something that needs to be tested as there are indications \citep[see][and references therein]{2020vandenBerg} that the emissivity of low-density open clusters frequently surpasses the emissivity of GCs, suggesting that dynamical destruction might be playing a role on the AB population too. It is also important to mention that a substantial fraction of the currently identified X-ray sources in GCs are unclassified sources and others are blends. Among those, there might be ABs, background/foreground sources, MSPs, qLMXBs, or even CVs. Therefore, when people consider the X-ray emissivity in a cluster, and despite the several strategies taken to remove nonmembers, there might still be a nonnegligible influence of background/foreground sources, which could lead to values that differ from the ``real" emissivity values for some clusters.

While strong dynamical interactions are able to help create the conditions in a binary to form a CV that otherwise would never have been formed, our results suggest that the detectable CV population is predominantly composed of CVs that have formed through the typical process of common-envelope evolution as correlations with cluster parameters are not observed.
It is important to stress that the stellar encounter rate is a parameter that varies throughout the cluster's evolution. Its value undergoes oscillations caused by episodes of core collapse during the cluster's life.
As a result, the present-day values of the stellar encounter rate ($\Gamma$) may not necessarily be comparable to values at earlier ages. 
Dynamics probably plays a role, not just in the formation of X-ray sources over the cluster's lifespan, but also in destroying their progenitors from primordial binaries.
 \cite{2019Belloni} found that a significant correlation exists between the proportion of primordial CV progenitors that are destroyed and the initial stellar encounter rate ($\Gamma_0$) of a cluster. In other words, a higher $\Gamma_0$ is associated with a more pronounced influence of dynamical interactions in the destruction of primordial CV progenitors. Simulations by \cite{2019Belloni} also show that GCs with similar present-day $\Gamma$ and similar $\Gamma_0$ can lead to a very different number of detectable CVs per unit of mass (by a factor $\gtrsim 10$). This can be explained as an interplay between dynamics destroying CV progenitors and change in the cluster global properties (e.g. the relaxation time, density, mass, core radius, and/or central velocity dispersion). However, detailed modeling to properly isolate the processes that change the cluster properties is needed to better understand this interplay.
\cite{2017Hong} also showed there is weak to no correlation between $\gamma$ and $N_{\rm CV}$ among CVs that have formed as a result of multiple weak and/or a few strong dynamical interactions, followed by common-envelope evolution. This is similar to what we observe. While very likely some detectable CVs have been formed via dynamical interactions, the current CV data suggest that the influence of the primordial population is more important than previously believed. Our findings indicate that the existing paradigm of CV formation largely being influenced by dynamical interactions might not be valid.

Interestingly, a correlation with $\Gamma$ has been observed for LMXBs and MSPs \citep[e.g.][]{2006Pooley,2013Bahramian,2023menezes}. This correlation can be understood based on the nature of the central energy source that drives GC evolution. \cite{2019Ye} have shown that there is an anticorrelation between the number of black holes (BHs) and MSPs in a GC. They found that if there is a large number of BHs in the cluster core, they will dominate the dynamics, preventing the concentration of neutron stars (NSs) in the central regions. This means that dynamical interactions would not contribute to forming binaries with either NSs or WDs. On the other hand, in clusters with less retained BHs, the NSs and mainly the most massive WDs (as they are more numerous) become the dominant source of energy \citep{Kremer2021}, and binaries hosting NSs, such as MSPs, and WDs can be more easily formed through dynamical interactions. However, even in such clusters, to form a pre-CV dynamically, the dynamical interaction has to involve a low-mass MS star, which seems to be less likely, as they are less numerous in the central parts \citep{Kremer2021}. Albeit further modeling is required to confirm this, we speculate that in clusters in which dynamics is dominated by WD subsystems, the number of dynamically formed binaries hosting WDs and low-mass MS stars is low, perhaps, only slightly higher than that in BH-dominated clusters, explaining in turn the lack of correlations we found here. The correlation with $\Gamma$ for LMXBs and MSPs, also suggests that, contrary to CVs, these binaries could have been formed more recently due to stellar interactions.

\begin{table}
\begin{center}
\begin{tabular}{ c  c  c  c }
 &  $\gamma$ vs $n_{\rm CV}$ & $\rho_c$ vs $n_{\rm CV}$ & $M$ vs $n_{\rm CV}$\\
 \hline
 \multicolumn{4}{c}{All Clusters}\\
 \hline
 A ; $\alpha$ ; C & 
$5.43\times 10^{-6}$  ; 1.58 ; 3.84 & 
 0.05 ; 0.34 ; 2.79 & 
$5.24\times 10^{8}$ ; $-1.53$; $3.39$\\ 
 $\chi^2$/dof  & 
97.35/18 & 
97.63/18  & 
95.36/18\\ 
$p$-value & 
 $6.76\times10^{-13}$ &
 $5.99\times 10^{-13}$ & 
 $1.55\times 10^{-12}$ \\ 
 \hline
 C & 
 4.19 & 
 4.19 & 
 4.19 \\
 $\chi^2$/dof & 
 103.70/20 & 
 103.70/20 & 
 103.70/20\\
 $p$-value & 
 $2.72 \times10^{-13}$ &
 $2.72 \times10^{-13}$ & 
 $2.72 \times10^{-13}$\\
  \hline
  A & 
  $2.11\times10^{-3}$ & 
  $8.21\times10^{-7}$ &  
  $2.09 \times10^{-6}$\\
  $\chi^2$/dof & 
  122.70/20  & 
  143.03/20 & 
  112.5/20  \\  
  $p$-value & 
  $1.11\times10^{-16}$ & 
  $<1\times10^{-17}$ & 
  $6.66\times 10^{-15}$ \\
 \hline
\multicolumn{4}{c}{Core-Collapsed Clusters}\\
\hline
 A ; $\alpha$ ; C &
 $3.40\times10^{6}$; $-1.46$ ; $10.9$ &
 $-5.39\times10^{3}$; $1.94\times10^{-3}$; $5.57\times10^{3}$ &
 $1.10\times10^{4}$; $-3.74\times10^{-4}$; $-1.09\times10^{4}$ \\ 
 $\chi^2$/dof & 
 2.10/3 & 
 15.01/3 & 
 25.63/3\\ 
 $p$-value & 
 0.55 & 
 $1.80\times10^{-3}$ & 
 $1.13\times10^{-5}$ \\ 
 \hline
 C & 
 23.35 & 
 23.35 & 
 23.35 \\ 
 $\chi^2$/dof & 
 26.01/5 & 
 26.01/5 & 
 26.01/5\\ 
 $p$-value &  
 $8.87\times10^{-5}$ & 
 $8.87\times10^{-5}$ & 
 $8.87\times10^{-5}$
 \\
  \hline 
  A & 
  $1.74\times10^{-3}$& 
  $7.20\times10^{-7}$ & 
  $5.38\times10^{-5}$\\ 
  $\chi^2$/dof & 
  39.51/5 & 
  48.31/5 & 
  35.13/5\\ 
  $p$-value & 
  $2.97\times10^{-7}$ & 
  $3.06\times10^{-9}$ & 
  $1.41\times10^{-6}$
  \\
\hline
\multicolumn{4}{c}{Non-Core-Collapsed Clusters}\\
\hline
 A ; $\alpha$ ; C &
$63.27$ ; $-0.90$ ; $2.70$ & 
$1.89\times10^{-6}$ ; 1.11 ; 3.44 &
$7.16\times10^{9}$ ; $-1.80$; $3.51$ \\ 
 $\chi^2$/dof & 
 54.73/12 & 
 58.23/12 & 
 57.40/12\\ 
 $p$-value & 
 $2.02\times10^{-7}$ & 
 $4.72\times10^{-8}$ &
$6.68\times10^{-8}$ \\ 
 \hline
 C & 
 3.82 & 
 3.82 & 
 3.82\\ 
 $\chi^2$/dof & 
 60.16/14 & 
 60.16/14 & 
 60.16/14\\ 
 $p$-value & 
 $1.10\times10^{-7}$ & 
 $1.10\times10^{-7}$ & 
 $1.10\times10^{-7}$ \\
  \hline
  A & 
  0.003& 
  $2.31\times10^{-5}$ & 
  $2.10\times10^{-5}$  \\ 
  $\chi^2$/dof & 
  81.27/14 & 
  81.91/14 &
  62.24/14\\ 
  $p$-value & 
  $1.63\times10^{-11}$ & 
  $1.24\times10^{-11}$ &
$4.71\times10^{-8}$ \\
\hline
\hline
\end{tabular}
\caption{\label{table2} Obtained values for the components in a relation of the types $Y= AX^\alpha$ + C, $Y=C$ and $Y=AX$ for different types of GCs. The value of $\chi^2$ and $p$-value for each fit is also given. All fits in the table were calculated including the data errors. This means that in the case of the constant relation, the values obtained differ from a simple average. Based on the fits, no relations are observed. Only in the case of $\gamma$ vs $n_{CV}$ for CC GCs there seems to be a power-law anticorrelation (see Sec.\ref{sec:discussion} for discussion).}
\end{center}
\end{table}

Our study is based on the most complete available sample of detectable CVs, which have information in more than one band or having spectra. Photometric surveys that are looking for CVs based on variability and colors (regardless of their X-ray emission) are on their way and will provide additional systems to the current sample of known CVs. Until then, we consider the sample presented in Table \ref{table1} is the most reliable sample of CVs to perform correlation studies. However, even when we consider the larger sample of CVs (see appendix \ref{appendix_ref} for details), which include those more ambiguous CVs --for example, detected only in X-rays-- we still do not observe a positive correlation (see Fig. \ref{figures_fit_increased} in appendix \ref{appendix_ref}). No clear trend was identified for any variable in this case. If we remove those clusters with one or two CVs in Table \ref{table1}, we still observe an anticorrelation between $n_{CVs}$ and $\gamma$ for CC GCs, and also, we observe a weak anticorrelation between $n_{CVs}$ and $M$ for CCs. This suggests that in CC GCs, the higher the dynamical interaction rate, the smaller the number of CVs, i.e., the dynamics contribute more to destroying CV progenitors than to creating CVs that would otherwise not become CVs.
If instead we remove clusters with one or two CVs from the more extended sample from the appendix \ref{appendix_ref}, our results are consistent with those when considering the filtered sample ($N_{CV}>2$) of Table \ref{table1}, i.e. we only observe an anticorrelation for CC GCs and $M$ above the $0.05$ significance level.
This clearly illustrates the importance of the number of GCs and CVs in the analysis, which is not surprising. In general, when adding all those more ``ambiguously classified" CVs (Table \ref{table_3}), our conclusions remain the same than when we use those CVs in table \ref{table1}, i.e. we do not see a positive correlation with $\gamma$ as mentioned in previous works.
If there is any significant (anti)correlation, it would most likely be an anticorrelation with $\gamma$ for CC GCs.

We emphasize that conclusions for the CV population based solely on their X-ray emission have to be taken with caution. This is because X-ray data alone may lead to misclassification. As discussed previously, CVs are relatively faint X-ray emitters, just as background/foreground sources, ABs, MSPs, or qLMXBs. Furthermore, radio studies have also reclassified some faint X-ray sources, initially classified as CVs, as binaries harboring BHs or pulsars \citep[e.g.,][]{2006Bogdanov,2015Miller-Jones}. This demonstrates the importance of combining multiwavelength studies to ``clean'' the sample of CVs and other close binaries as much as possible. In the last years, UV observations have shown to be extremely useful to detect very faint CVs or systems in very crowded areas. However, UV observations are not always possible to carry out due to large stellar extinction to many clusters, including those with very large numbers of X-ray sources such as Terzan 5. In this regard, new facilities, such as the James Webb Space Telescope (JWST), would allow us to observe the infrared counterparts, providing a more robust classification. 

\cite{2019Belloni} determined that the detectable\footnote{The criteria to be selected as observable CVs used by these authors are consistent with the characteristics of the observed population in the clusters 47 Tuc, $\omega$ Cen, NGC 6397, and NGC 6752.} CVs in GCs correspond to 2-4\% of all the existing CVs in GCs (the percentage depending on the common-envelope phase efficiency). If we assume these models are correct, this means that in order to reduce the current uncertainties in the observations, and to allow the identification of a clear trend, we would need to detect a number of CVs that is larger by a factor of several tens (25-50, if we want to identify all CVs) than the current identified sample per cluster. Therefore, unless we are able to do so with telescopes such as the JWST or the Rubin Observatory\footnote{The Rubin Observatory will not resolve the cores of GCs as has been done by HST or MUSE. However, its wide field of view, its multicolor photometry, and the 10 yr duration of Rubin LSST, would allow us to identify CVs in different parts of GCs through their variability \citep{2023Hambleton}, including those residing in the outskirts, where very few have been identified so far.}, our conclusions will likely remain the same, i.e., we will not be able to identify a possible relation. However, the identification of a trend with smaller CV samples would impose constrains to models, for instance, on the common-envelope phase efficiency. It would be interesting to further investigate that.


\begin{acknowledgments}
We thank the referee for their comments which helped to improve this manuscript. L.R.S thanks Dr. Yuri Cavecchi and  Prof. Thomas Maccarone for useful discussions.
D.B. acknowledges financial support from {FONDECYT} grant number {3220167}. 
\end{acknowledgments}

\software{astropy \citep{2013A&A...558A..33A,2018AJ....156..123A} }


\bibliography{bibliography}{}
\bibliographystyle{aasjournal}

\appendix
\section{Appendix: List of detectable CVs in Galactic Globular clusters}
\label{appendix_ref}

\begin{figure*}
\begin{center}
\includegraphics[width=0.99\linewidth]{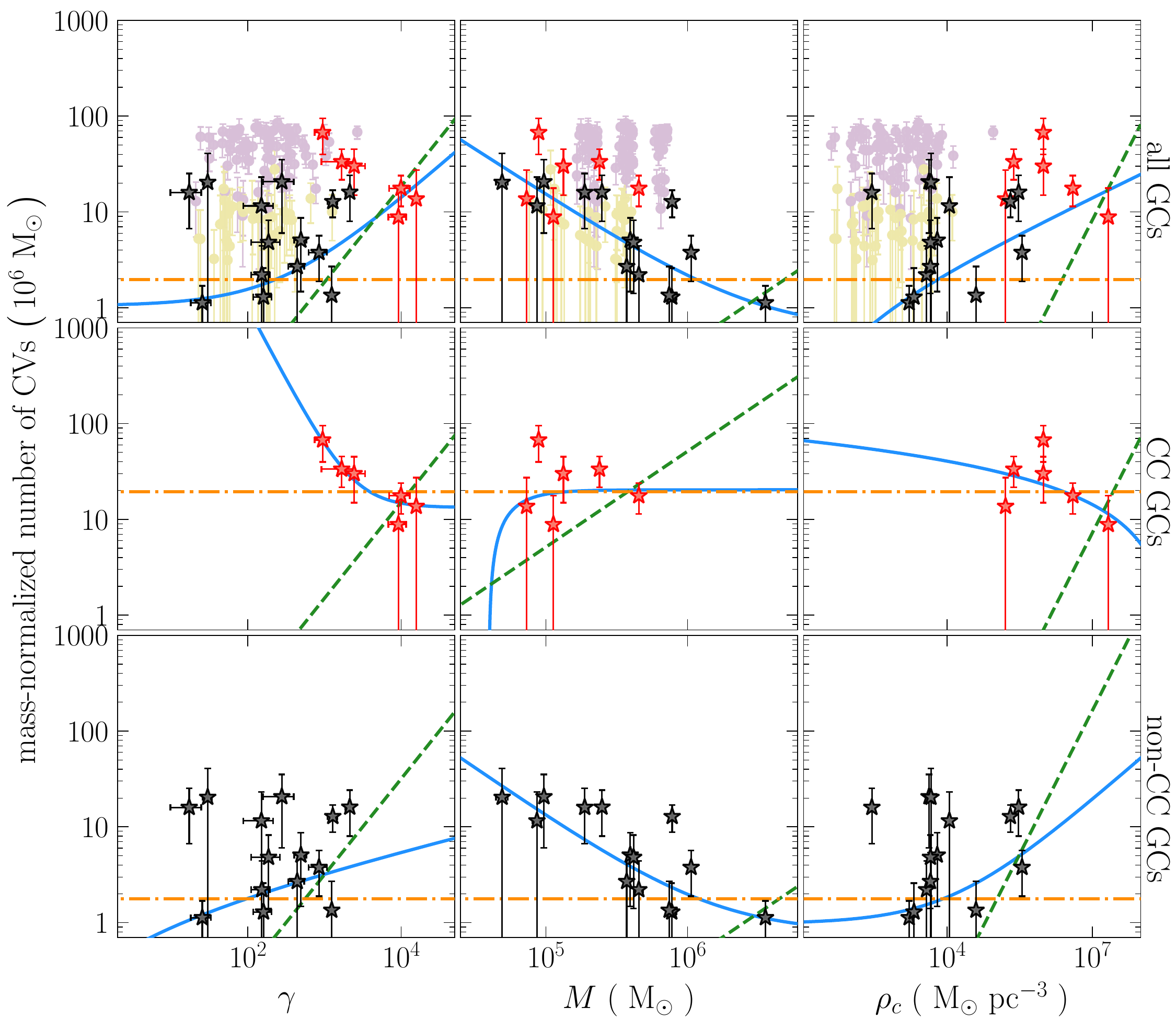}
\end{center}
\caption{ 
Relation between $n_{CV}$ vs cluster variables
but for ``bright" CVs in the sample of Table \ref{table1}, i.e. those brighter than optical absolute magnitude 9.
Symbols, colors and lines have the same meaning as Figure \ref{figures_fit}.
}
\label{figures_fit_bright}
\end{figure*}

 Below is the literature from which we have built the sample of detectable CVs for this paper and reported in Table~\ref{table1}. Figure \ref{figures_fit_bright} shows the relations between $n_{CV}$ vs $\gamma$, $M$ and $\rho_c$ for the systems in Table \ref{table1} which are brighter than optical absolute magnitude 9. Systems with an * symbol were not counted in Table \ref{table1} due to large uncertainty in their classification, often caused by having emission in one single band (X-Ray, UV, optical or IR), having more than one plausible counterpart within the X-ray error circle, or characteristics being well compatible with other types of sources, but are listed here for completeness. The references given right after the lists of IDs in each cluster refer only to the systems marked with an *. We added the detectable CVs from Table \ref{table1} to the systems marked with a * to form Table~\ref{table_3}. The results of  the analysis described in Section \ref{Data_analysis} but performed on the sample of Table \ref{table_3} are shown in Figure~\ref{table_3}. \\

References for the systems used in Table \ref{table1}, references therein are often not cited here as it is beyond the scope of this paper to provide the history for each individual source. 
NGC 104 (47 Tuc): \cite{2018RiveraS}; 
NGC 6397: \cite{2010cohn, 2010Bogdanov,2017Dieball};  
NGC 6752: \cite{2014Forestell, 2017ALugger,2021Cohn}; 
NGC 288: \cite{2006Kong}; 
NGC 6121 (M4): \cite{2004Bassa, 2023Lugger}; 
NGC 6809 (M55): \cite{2008Bassa}; 
NGC 6838 (M71): \cite{2010Huang}; 
NGC 6218 (M12): \cite{2009Lu,2019Gottgens};
NGC 6388: \cite{2012Maxwell}; 
NGC 7099 (M30): \cite{2007Lugger,2020Zhao,2022Mansfield,2019Gottgens};
NGC 6624: \cite{1999Deutsch,1996Shara}; 
NGC 6093 (M80):  \cite{2003Heinke,2019Gottgens,2008Pietrukowicz,2005Shara}; 
NGC 6205 (M13): \cite{2011Servillat}; 
NGC 7078 (M15): \cite{2007Dieball,Hannikainen2005}; 
NGC 5272 (M3): \cite{2019Zhao} ; 
NGC 5139 ($\omega$ Cen): \cite{2013Cool,2018Henleywillis,2019Gottgens}; 
NGC 6656 (M22): \cite{2005Pietrukowicz,2019Gottgens,2019Gottgensa}, NGC 6681 (M70): \cite{2019Gottgens}; NGC 5904 (M5): \cite{2011Hourihane}; NGC 6402 (M14): \cite{1964sawyer,2024Zhao}, NGC 2808:  \cite{2008Servillat}. \\



Non-Core-Collapsed:
\begin{itemize}[label=, left=0pt, itemsep=3pt, topsep=5pt]
    \item NGC 104 (47 Tuc): W1, W2, W8, W15, W16, W23, W24, W25, W27, W30, W33, W35, W36, W44, W45, W49, W51, W53, W55, W56, W75, W98, W100-a/100-b, W120, W122, W140, W187, W200, W202, W223, W229, W235, W256, W292, W295, W299, W302, W304, W309, W311, W324, W329, W335.\\
    W20$^*$\footnote{Multiwavelength analysis suggests the object W32$^{*}$ listed in \cite{2023Bao} as CV candidate, is instead an AB. We have also excluded the AM~CVn candidate in 47 Tuc and in NGC 1851.},
    No.263$^{*}$, No.$267^{*}$, No.$320^{*}$, No.$341^{*}$, No.$376^{*}$, No.$378^{*}$, No.$395^{*}$, No.$486^{*}$, No.$501^{*}$, No.$522^{*}$. From \cite{2022Saeedi,2023Bao}
    \item NGC 288: CX6*, CX13$^{*}$, CX17*, CX20*, CX24, CX27*. From \cite{2006Kong}.
    \item NGC 2808: Chandra (C) 2$^*$, C3$^*$, C4$^*$, C5$^*$, C7, C8$^*$, C9$^*$, C10$^*$,  C11$^*$, C12$^*$, C13$^*$, C14$^*$, C15$^*$, C16$^*$, C17$^*$, XMM-Newton source C5$^*$. From \cite{2008Servillat}
    \item NGC 5139 ($\omega$ Cen): 12a, 13a, 13c, 13f, 21b, 22c, 23b, 24c, 31a, 32a, 33d, 33e, 33j, 33m, 41a, 41d, 43h, 44d, 51d, 52d, 54b, 54e, 54h, 21c$^{*}$, 21d$^{*}$, 44c$^{*}$. From \cite{2018Henleywillis}
    \item NGC 5272 (M3): CX1 (1E1339), CX6, CX7$^{*}$, CX8, CX12$^{*}$, CX13$^{*}$, CX16$^{*}$. From \cite{2019Zhao}.
    \item NGC 5904 (M5): V101.
    \item NGC 6093 (M80): DN1 (CX7), CX1 (T Sco), CX2$^{*}$, CX3, CX4$^{*}$, CX5 (DN2), CX13$^{*}$, CX15$^{*}$, CX16$^{*}$, CX17$^{*}$. From \cite{2003Heinke,2010Dieball}
    \item NGC 6121 (M4): CX2, CX4, CX76, CX81, CX101$^{*}$. From \cite{2023Lugger}.
    \item NGC 6205 (M13): X6.
    \item NGC 6218 (M12): CX1, CX2$^{*}$. From \cite{2009Lu}
    \item NGC 6304: CXOU J171431.86-292745.5$^*$. From \cite{2009Guillot}
    \item NGC 6341 (M92): CX1$^*$, CX2$^*$, CX3$^*$, CX4$^*$, CX5$^*$. From \cite{2011Lu92}
    \item NGC 6366: CX1$^{*}$. From \cite{2008Bassa}.
    \item NGC 6388: CX4$^*$, CX5, CX7, CX9, CX11, CX12. From \cite{2012Maxwell}
    \item NGC 6402 (M14): Oph 1938.
    \item NGC 6440: CX4$^{*}$, CX6$^{*}$, CX7$^{*}$, CX10$^{*}$, CX12$^{*}$, CX13$^{*}$, CX24$^{*}$. From \cite{2002pooley}. 
    \item NGC 6553: Sgr 1943$^{*}$. From \cite{1949Mayall} and \cite{2022Bond}, found at the edge of the cluster, so its membership is unclear.
    \item NGC 6626 (M28): Seq. 1*/ Seq. 4$^{*}$, Seq. 2$^{*}$, Seq. 3$^{*}$, Seq. 7$^{*}$ (Seq. 1 and Seq. 4 seem to be associated to the same X-ray source). From \cite{2024Bao}
    \item NGC 6652: C$^*$ (V12), E$^*$. From \cite{2012Stacey}
    \item NGC 6656 (M22): CV1\footnote{It is the same object listed in \cite{2005Pietrukowicz} as CV1}, CV2, Novae1. 
    \item NGC 6715 (M54): ID 1$^{*}$, ID 2$^{*}$, ID 3$^{*}$, ID 4$^{*}$, ID 5$^{*}$, ID 6$^{*}$, ID 7$^{*}$. From \cite{2006Ramsay}.
    \item NGC 6809 (M55): CX1, CX2, CX20, CX21$^{*}$, CX24. From \cite{2008Bassa}.
    \item NGC 6838 (M71): S29, S54$^{*}$, S41$^{*}$, S25$^{*}$, S21$^{*}$, S14$^{*}$, S10$^{*}$, S05$^{*}$. From \cite{2010Huang}.
    \item Djorg 2 (ESO 456-SC38): OGLE-UCXB-01$^{*}$. From \cite{2020Bahramian}.
    \item Terzan 5 (Terzan 11): Seq. 1$^{*}$, Seq. 2$^{*}$, Seq. 3$^{*}$, Seq. 4$^{*}$, Seq. 5$^{*}$, Seq. 7$^{*}$, Seq. 8$^{*}$. From \cite{2024Bao}.
\end{itemize}

Core-Collapsed:\\
\begin{itemize}[label=, left=0pt, itemsep=3pt, topsep=5pt]
    \item NGC 6266 (M62): s06$^{*}$, s11$^{*}$, s16$^{*}$, s18$^{*}$, s19$^{*}$, s24$^{*}$, s25$^{*}$, s28$^{*}$, s29$^{*}$, s30$^{*}$, s32$^{*}$, s34$^{*}$. From \cite{2020Oh}.
    \item Terzan 1 (HP 2): Terzan 1 CX1$^{*}$. From \cite{2020Bahramian}.
    \item NGC 6397: U7, U10, U11, U13, U17, U19, U21, U22, U23, U25, U31, U60, U61, U80, U83, Seq. 2$^{*}$. From \cite{2024Bao}.
    \item NGC 6624: Star A, DN1.
    \item NGC 6681 (M70): CV1.
    \item NGC 6752: CX1, CX2, CX4, CX5*, CX6, CX7, CX9, CX13, CX21, CX23, CX24, CX25, CX29, CX32, CX35, CX36, CX41, CX49. From \cite{2024Bao}
    \item NGC 7078 (M15): CV1 (HCV 2005-A), HCV 2005-B$^{*}$, M15-C, V5$^{*}$, V6$^{*}$, V7, V11, V15, V20$^{*}$, V26$^{*}$, V37$^{*}$, V38$^{*}$, V39, V40, V41. From \cite{Hannikainen2005,2007Dieball}\footnote{ Other possible CV candidates are not explicitly labeled in \cite{2007Dieball}.} 
    \item NGC 7099 (M30): A2$^{*}$, A3, B, C, 8, 9, 12, W15, W19, W21.\footnote{Object A2 was classified as an RS CVn? by \cite{2020Zhao}, but more recently as a CV by \cite{2022Mansfield} which is consistent with the classification given previously by \cite{2007Lugger}. Object M30-V4 is a confirmed dwarf nova, but it seems to be a foreground object \citep{2008Pietrukowicz}.}
\end{itemize}

\begin{table}[h]
\begin{center}
\begin{tabular}{ |c|c| }
\hline
Cluster & Total Detections\\
\hline
\multicolumn{2}{c}{Non-core-collapsed}\\
\hline
NGC 104 (47 Tuc) & 54  \\ 
NGC 288 & 6  \\ 
NGC 2808 & 16 \\
NGC 5139 ($\omega$ Cen) & 26  \\ 
NGC 5272 (M3) & 7\\
NGC 5904 (M5) & 1 \\
NGC 6093 (M80) & 10 \\
NGC 6121 (M4)  & 5\\
NGC 6205 (M13)  & 1 \\
NGC 6218 (M12)  & 2 \\ 
NGC 6304  & 1\\
NGC 6341 (M92) & 5 \\
NGC 6366 & 1\\
NGC 6388 & 6 \\
NGC 6402 (M14) & 1\\
NGC 6440 &  7\\
NGC 6553 & 1\\
NGC 6626 (M28) & 4\\
NGC 6652  & 2\\
NGC 6656 (M22) & 3 \\
NGC 6715 (M54) & 7 \\
NGC 6809 (M55) & 5 \\
NGC 6838 (M71) & 8\\
Djorg 2  & 1 \\
Terzan 5  & 7 \\

\hline
\multicolumn{2}{c}{Core-collapsed }\\
\hline
NGC 6266 (M62) & 12 \\
Terzan 1 & 1 \\
NGC 6397 & 16 \\
NGC 6624 & 2 \\
NGC 6681 (M70) & 1 \\
NGC 6752 & 18 \\
NGC 7078 (M15) & 15 \\
NGC 7099 (M30) & 10\\
 \hline
\end{tabular}
\caption{A summary of the number of systems that have been classified in the literature as CVs or potential CVs. The total number per cluster include all the systems mentioned in the Appendix. Systems listed in Table \ref{table1} are included here.}
\label{table_3}
\end{center}
\end{table}

\begin{figure*}
\begin{center}
\includegraphics[width=0.99\linewidth]{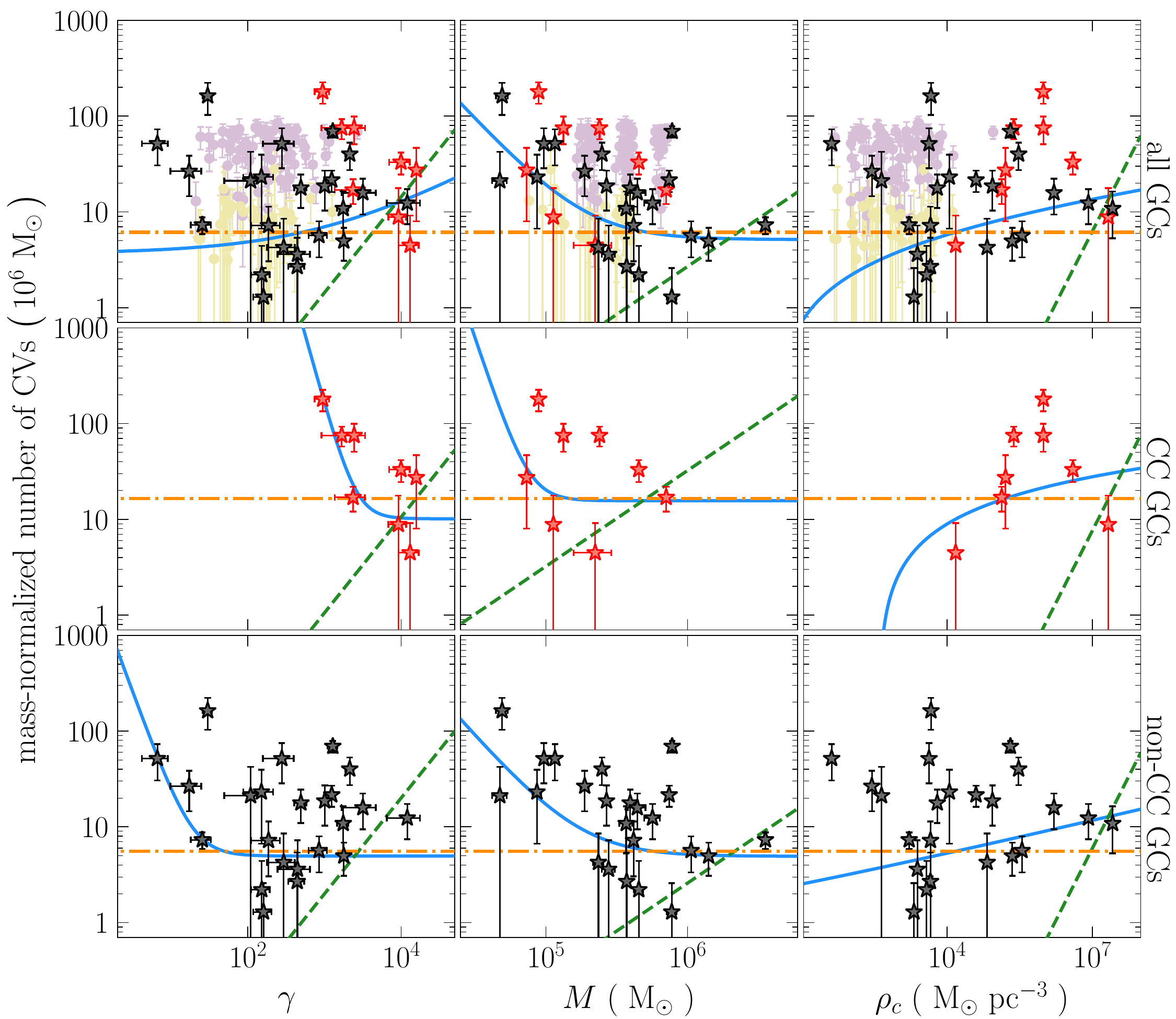}
\end{center}
\caption{
Relation between $n_{CV}$ vs cluster variables
but for all the CVs listed in the appendix, regardless of whether they have been detected in just one or more bands. 
Symbols, colors and lines have the same meaning as Figure \ref{figures_fit}.}
\label{figures_fit_increased}
\end{figure*}

\end{document}